\documentclass[namedreferences]{SolarPhysics}
\usepackage{epsfig}

\usepackage{graphicx} 

\begin{document}
\begin{article}
\begin{opening}
\title{COMPUTING NONLINEAR FORCE-FREE CORONAL MAGNETIC FIELDS IN SPHERICAL GEOMETRY}
\author{T. \surname{WIEGELMANN}\email{wiegelmann@mps.mpg.de}}
\institute{Max-Planck-Institut f\"ur Sonnensystemforschung,
Max-Planck-Strasse 2, 37191 Katlenburg-Lindau, Germany}


\runningtitle{Spherical force-free fields}
\runningauthor{Wiegelmann}
\date{Received ; accepted }

\begin{abstract}
 We describe a newly developed code for the extrapolation of nonlinear
 force-free coronal magnetic fields in spherical coordinates.
 The program uses measured vector magnetograms on
 the solar photosphere as input and solves the force-free equations in
 the solar corona. The method is based on an optimization principle
 and the heritage of the newly developed code is a corresponding method in Cartesian
 geometry. We test the newly developed code with the help of a semi-analytic
 solution and rate the quality of our reconstruction qualitatively by
 magnetic field line plots and quantitatively with a number of comparison metrics.
 We find that we can reconstruct the original test field with high accuracy.
 The method is fast if the computation is limited to low co-latitudes
 (say $30^\circ \leq \theta \leq 150^\circ$), but becomes significantly slower
 if the polar regions are included.
\end{abstract}
\end{opening}

\section{Introduction}
 The solar corona is dominated by the coronal magnetic field because
 the magnetic pressure is several orders of magnitude higher than
 the plasma pressure. Knowledge regarding the coronal magnetic field
 is therefore important to understand the structure of the coronal
 plasma and to get insights regarding dynamical processes like
 flares and coronal mass ejections. The direct measurement of
 magnetic fields is very difficult and such measurements are only
 occasional available, see, e.g., \inlinecite{lin:etal04}. Well established
 are measurements of the photospheric magnetic field with the help of
 line-of-sight magnetographs (e.g., SOHO/MDI, NSO/Kitt Peak) or
 vector magnetographs (e.g., currently with the Solar Flare Telescope/NAOJ, Imaging vector
 magnetograph/MEES Observatory, and in near future with SOLIS/NSO, Hinode, SDO/HMI).
 These photospheric fields can be extrapolated into the solar corona by
 making suitable model assumptions. The simplest approach is to assume
 current-free potential fields, which can be computed from the photospheric line-of-sight
 magnetic field alone. Such source surface potential field models
 \cite{schatten69,neugebauer:etal02,schrijver:etal03} give already some
 insights about the coronal magnetic field structure, e.g., regarding the
 location of coronal holes and active regions. They cannot, however,
 be used to estimate the magnetic energy
 and helicity of the corona, in particular not the free energy available for
 eruptive phenomena. For reliable estimates of these quantities
 currents have to be included. To lowest order the
 effects of plasma pressure and gravity can be neglected and we can assume
 that the currents are parallel to the magnetic field,
 the force-free assumption. A popular simplification of force-free fields are linear force-free
 fields \cite{chiu:etal77,seehafer78} where the electric current flow is parallel to
 the magnetic field with a global constant of proportionality $\alpha$. This approach
 is in particular popular because linear force-free models require only line-of-sight
 magnetograms as input and contain only a single free parameter $\alpha$, which can be specified
 by comparing magnetic field line plots with coronal images \cite{carcedo:etal03,marsch:etal04}.
 In general $\alpha$ changes in space and taking this into account requires the
 use of nonlinear force-free models. A comparison of observationally inferred 3D
 magnetic field structures in a newly developed active regions \cite{solanki:etal03} with
 different extrapolated field models by \inlinecite{wiegelmann:etal05} revealed that
 linear force-free fields are better than potential fields, but nonlinear force-free
 models are even more accurate. The computation of nonlinear force-free fields
 is more challenging, however for several reasons. Mathematically problems regarding
 the existence and uniqueness for various boundary value problems dealing with
 nonlinear force-free fields are still open, see \inlinecite{amari:etal06} for details.
 Another issue is the numerical analysis for a given boundary value problem. An additional
 complication is to derived the required boundary data from photospheric vector magnetic field
 measurements. High noise in the transversal components of the measured field vector, ambiguities
 regarding the field direction and nonmagnetic forces in the photosphere complicate
 the task to derive suitable boundary conditions from measured data.

 Different approaches have been proposed for the nonlinear force-free
 extrapolation of vector magnetograms:
\begin{itemize}
 \item Upward integration method, (e.g., \opencite{wu:etal90},
 \opencite{cuperman:etal91}, \opencite{demoulin:etal92}, \opencite{amari:etal97}).
 \item Grad-Rubin-like method, (e.g., \opencite{grad:etal58},
 \opencite{sakurai81}, \opencite{amari:etal97}, \opencite{amari:etal99},
 \opencite{wheatland04}, \opencite{amari:etal06}).
 \item Different MHD relaxation methods, (e.g., \opencite{mikic:etal94},
 \opencite{roumeliotis96}, \opencite{valori:etal05}).
 \item Green's function like method, (e.g., \opencite{yan:etal00}, \opencite{yan:etal06}).
 \item Optimization method, (e.g., \opencite{wheatland:etal00}, \opencite{wiegelmann04}).
\end{itemize}
For a more complete review on existing methods for computing nonlinear force-free
 coronal magnetic fields see the review papers by
 \inlinecite{amari:etal97} and \inlinecite{schrijver:etal06}.
 The Grad-Rubin method as described in \inlinecite{amari:etal97} and
 \inlinecite{amari:etal99}
  has been applied to investigate  particular active regions in
  \inlinecite{bleybel:etal02} and a comparison of the extrapolated field with
  2D projections of plasma structures as seen in H$\alpha$, EUV and X-ray has been done in
  \inlinecite{regnier:etal02} and \inlinecite{regnier:etal04}. The optimization code in the
  implementation of \inlinecite{wiegelmann04} has been applied to an active region
  in \inlinecite{wiegelmann:etal05a} and compared with H$\alpha$ images.
  \inlinecite{wiegelmann:etal06a} investigated the possibility to use magnetic
  field extrapolations to improve the stereoscopic 3D reconstruction from
  coronal images observed from two viewpoints.

 Recently \inlinecite{schrijver:etal06} compared the performance of six different
 Cartesian nonlinear force-free extrapolation codes in a blind algorithm test.
 All algorithms yield nonlinear force-free fields that
 agree well with the reference field in the deep interior of the volume, where the field
 and electrical currents are strongest. The optimization approach successfully
 reproduced also the weak field regions and compute the magnetic energy content
 correctly with an accuracy of $2 \%$.
 In a coordinated study
 \inlinecite{amari:etal06} obtained an accuracy of somewhat better than $2 \%$.

 The good performance of the optimization method encourages
 us to develop a spherical version of the optimization code.
 The required full disk vectormagnetograms will become available
 soon  (e.g., from SOLIS).  The heritage of the newly developed
 code is a  Cartesian force-free optimization method as implemented by
 \inlinecite{wiegelmann04}. We outline the paper as follows. In Section \ref{magnetic_modelling}
 we describe our newly developed algorithm. Section \ref{tests} contains a semi-analytic
 test case and the setup of computations to check the accuracy and performance of our code.
 We introduce figures of merit to rate the quality of our reconstruction in Section
 \ref{merit} and present the results of our test runs in Section \ref{results}.
 Finally we draw conclusions
 in Section \ref{conclusions} and give an outlook for future work.

\section{Method}
\label{magnetic_modelling}
Force-free magnetic fields have to obey the equations
\begin{eqnarray}
(\nabla \times {\bf B }) \times{\bf B} & = & {\bf 0},
\label{forcefree} \\
\nabla \cdot{\bf B}    & = &         0.
\label{solenoidal-ff}
\end{eqnarray}

We solve  Equations (\ref{forcefree}) and (\ref{solenoidal-ff}) with the help of an
optimization principle  as proposed by \inlinecite{wheatland:etal00} and generalized
by \inlinecite{wiegelmann04}. Until now the method has been implemented in Cartesian
geometry.

Here we define a functional in spherical geometry:
\begin{equation}
L=\int_{V}   \left[B^{-2} \, |(\nabla \times {\bf B}) \times {\bf
B}|^2 +|\nabla \cdot {\bf B}|^2\right] \, r^2 \, \sin \theta \, dr \, d \theta \, d \phi
\label{defL1},
\end{equation}
 It is obvious that the
force-free Equations (\ref{forcefree}-\ref{solenoidal-ff}) are fulfilled when $L$ equals
zero. We normalize the magnetic field with the average radial magnetic
 field on the photosphere and the length scale with a solar radius.

 The functional (\ref{defL1}) can be numerically minimized
 with the help of the iteration equations:
\begin{equation}
\frac{\partial {\bf B}}{\partial t} =\mu {\bf F}
\label{iterateB}.
\end{equation}
where $\mu$ is a positive constant and
\begin{eqnarray}
{\bf F} & =& \nabla \times ({\bf \Omega}_a \times {\bf B} )
- {\bf \Omega}_a \times (\nabla \times {\bf B})  \nonumber\\
& & +\nabla({\bf \Omega}_b \cdot {\bf B})-  {\bf \Omega}_b(\nabla \cdot {\bf B})
+({\bf \Omega}_a^2 + {\bf \Omega}_b^2)\; {\bf B}
\end{eqnarray}
with
\begin{eqnarray}
{\bf \Omega}_a &=&
B^{-2} \;\left[(\nabla \times {\bf B})\times {\bf B} \right] \\
{\bf \Omega}_b &=& B^{-2} \;\left[(\nabla \cdot {\bf B}) \; {\bf B} \right].
\label{defomega}
\end{eqnarray}
The theoretical deviation of the iterative Equation (\ref{iterateB}) as outlined
 by \inlinecite{wheatland:etal00}
 does not depend on the use of a specific coordinate system. Previous numerical
implementations of this method have been done to our knowledge only in Cartesian
geometry, however. Here we describe a newly developed implementation in spherical
geometry.

\subsection{Implementation}
 We use a spherical grid $r, \theta, \phi$ with $n_r, \,
n_{\theta}, \, n_{\phi}$ grid points in radial direction, latitude
\footnote{$\theta$ corresponds to the co-latitude, with $\theta=0^\circ$ and
 $\theta=180^\circ$ at the south and north poles, respectively.}
and longitude,
respectively. Here we intend to compute the whole sphere $r=1 R_{\rm s} \dots 2.57 R_{\rm s}$,
$\, \theta=0^\circ \dots 180^\circ$, $\, \phi=0^\circ \dots 360^\circ$, but in principle one could
limit the method also to parts of a sphere. To avoid the mathematical
singularities at the poles, we do not use grid points exactly at the south and north
pole, but half a grid point apart $\theta_{\rm min}=\frac{d \theta}{2}$ and
$\theta_{\rm max}=180^\circ-\frac{d \theta}{2}$.

The method works as follows:
\begin{enumerate}
\item We compute the initial source surface potential field in the computational
domain from $B_r$ on the photosphere at $r= 1 R_S$ .
\item We replace $B_{\theta}$ and $B_{\phi}$ at the bottom photospheric boundary at $r= 1 R_S$ by the measured
vector magnetogram.
 \footnote{It is as well possible to replace
only parts of the photosphere (a limited region in $\theta$ and $\phi$ direction) and
to restrict the nonlinear force-free computation onto this region. This is in particular
necessary if the observed photospheric vector magnetogram is only available for parts
of the photosphere. In such cases the global magnetic field is basically a potential
field and only locally (say in active regions) a nonlinear force-free field. For such
limited regions in $\theta$ and $\phi$ we encounter the same problem as in Cartesian codes,
that the lateral boundary conditions are unknown. One possibility is to describe the lateral
boundaries with the help of a global potential field. The assumption of a potential field outside
the computational domain restricts currents to the active region, but
non current carrying field lines can leave the computational box.
At the interface between the potential and non-potential field, a boundary layer as described
in \inlinecite{wiegelmann04} for the Cartesian implementation of our code can be used.
Full spherical force-free fields certainly do not have lateral boundaries and
if one is interested in the details of interaction of two far apart active regions, it might
be better to compute first a global low resolution
force-free field (e.g., in future with SOLIS data) and then compute the
field in the active regions with higher resolution (e.g., with Hinode data).
Such an approach would allow also current carrying
field lines to connect the two active regions, which might be important
for the initiation of CMEs.}
 The  outer radial boundary is unchanged from the initial potential field
model. For the purpose of code testing we try additional other boundary conditions,
see Section \ref{boundary}.
\item We iterate for a force-free magnetic field in the computational box by minimizing
the functional $L$ of Equation (\ref{defL1}) by applying Equation (\ref{iterateB}).
\item The continuous form of Equation (\ref{iterateB}) ensures a monotonically
decreasing functional $L$. For finite time steps, this is also ensured if
the iteration time step $dt$ is
sufficiently small. If $L(t+dt) \geq L(t)$ this step is rejected and we repeat
this step with $dt$ reduced by a factor of two.
\item After each successful iteration step we increase $dt$ by a factor of $1.01$ to
ensure a time step as large as possible within the stability criteria. This ensures
an iteration time step close to its optimum.

\item The iteration stops if $dt$ becomes too small. As stopping criteria we use
 $dt \leq 10^{-9}$.
\end{enumerate}
\section{Test Case}
\label{tests}
\begin{figure}
\includegraphics[bb=0 0 400 400,clip,height=6cm,width=12cm]{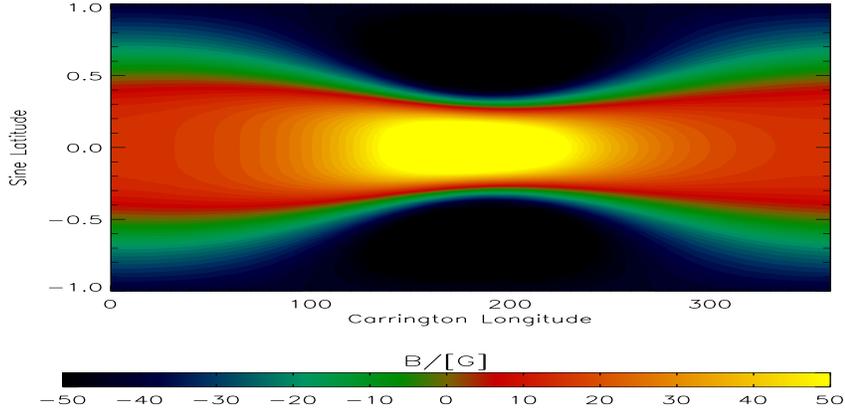}
\caption{The synoptic map shows $B_r$ on the photosphere.
 We project the map onto the solar surface in Figure \ref{fig2} with
 the disk center at Carrington Longitude $180^\circ$.}
 \label{fig1}
\end{figure}
\begin{figure}
\mbox{
\includegraphics[bb=0 20 400 430,clip,height=6cm,width=7cm]{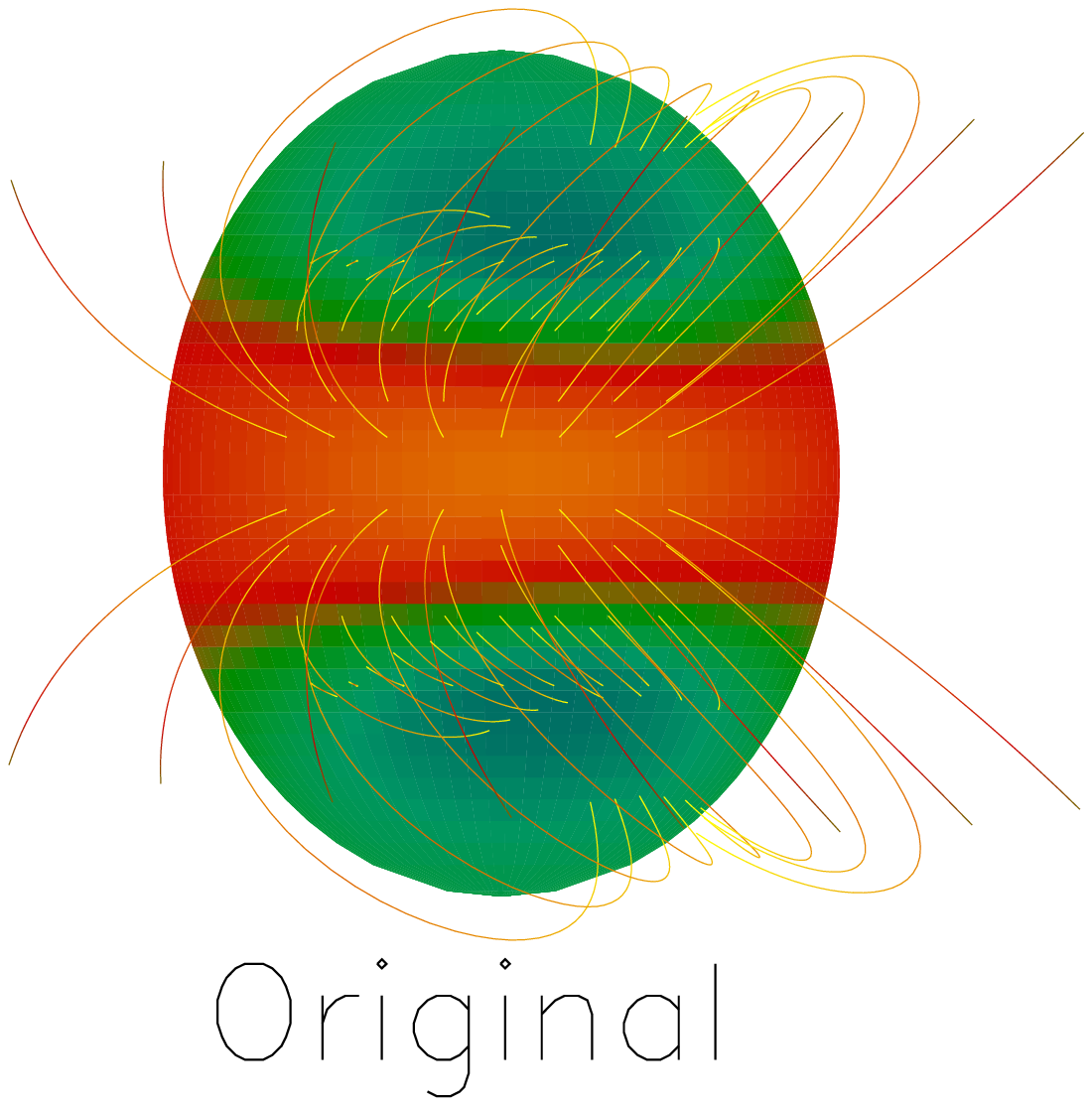}
\includegraphics[bb=0 20 400 430,clip,height=6cm,width=7cm]{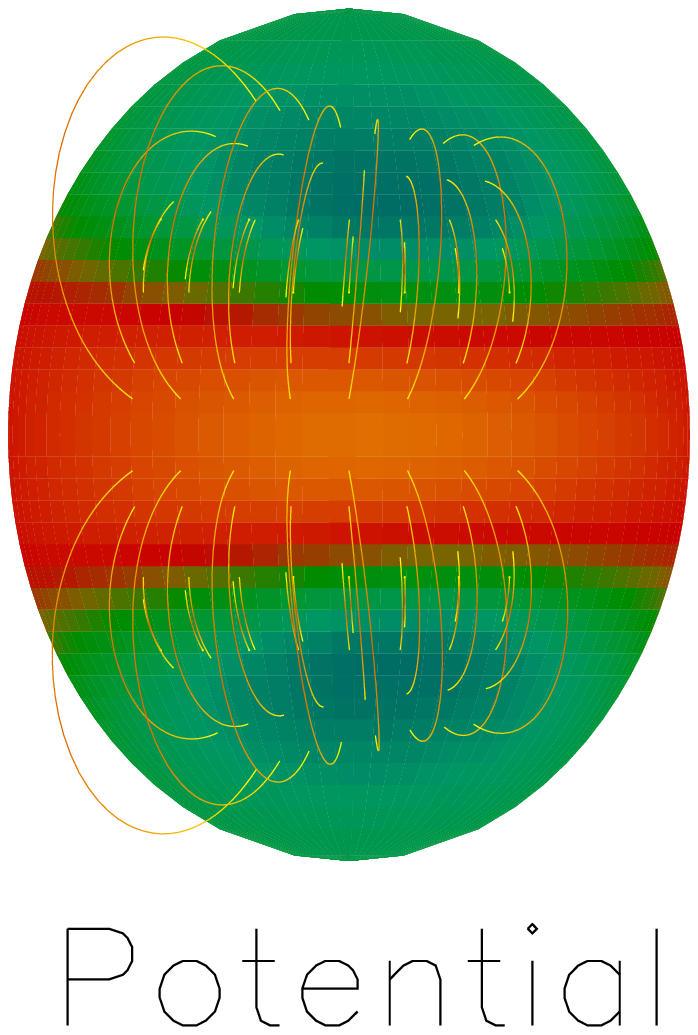}}
\mbox{
\includegraphics[bb=0 20 400 430,clip,height=6cm,width=7cm]{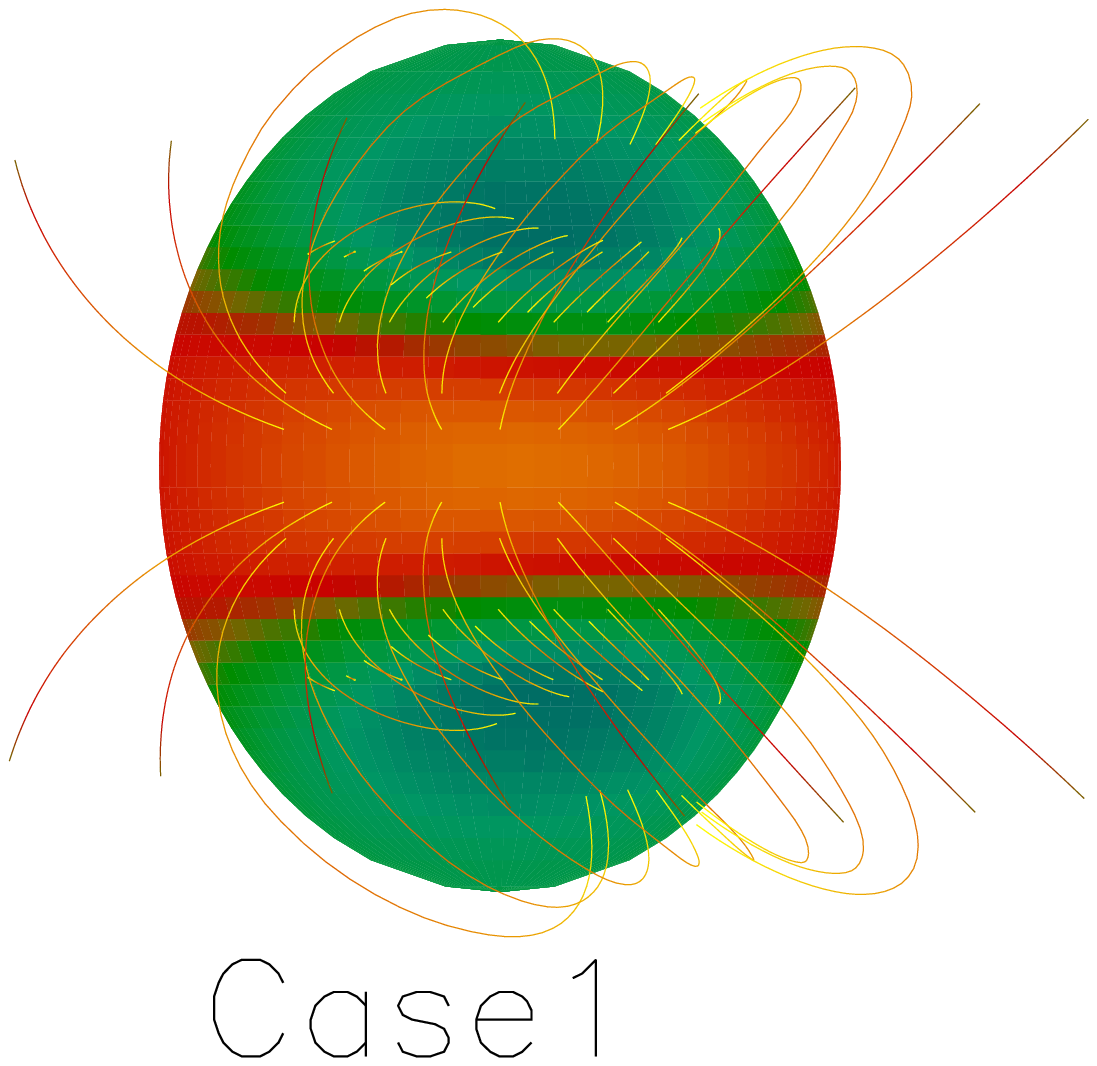}
\includegraphics[bb=0 20 400 430,clip,height=6cm,width=7cm]{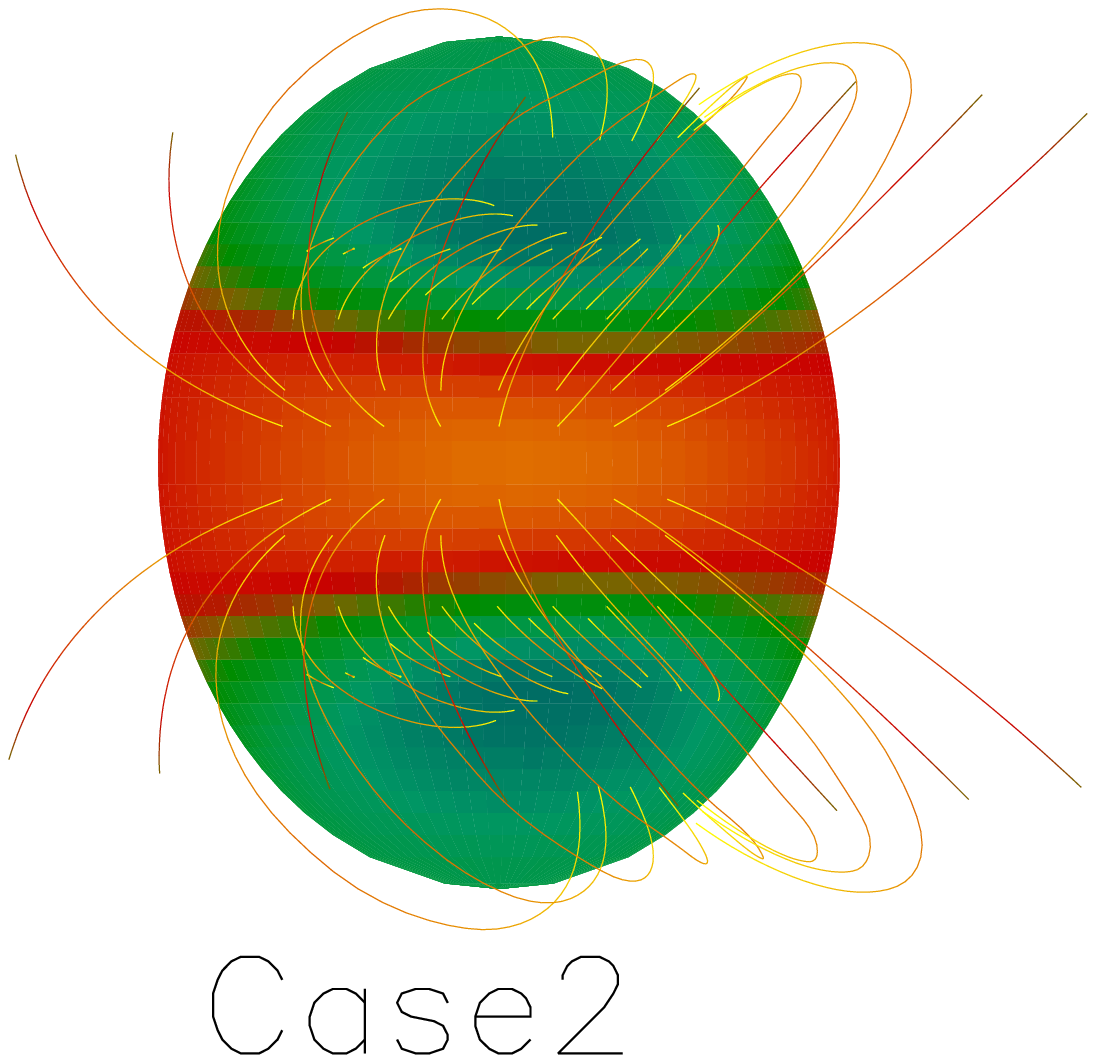}}
\mbox{
\includegraphics[bb=0 20 400 430,clip,height=6cm,width=7cm]{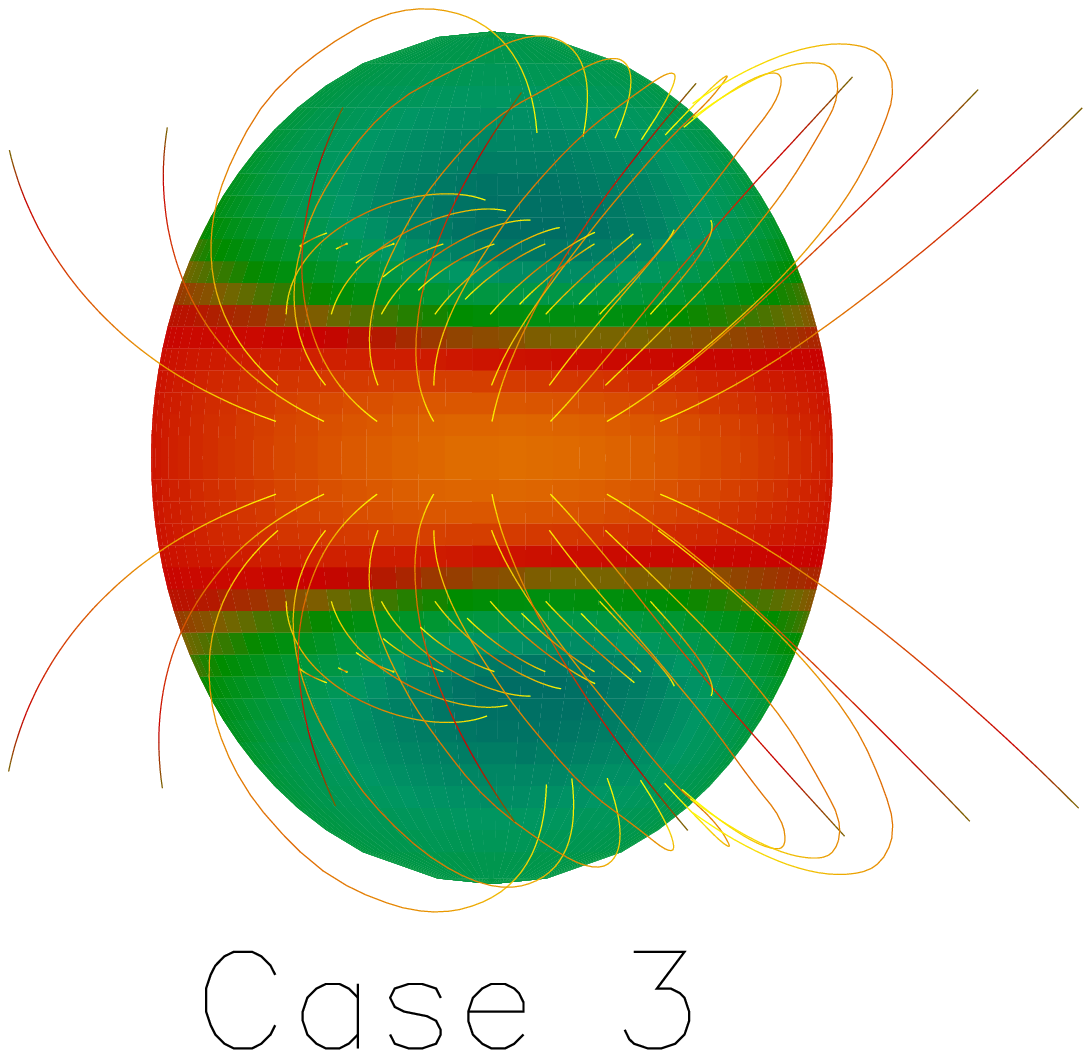}
\includegraphics[bb=0 20 400 430,clip,height=6cm,width=7cm]{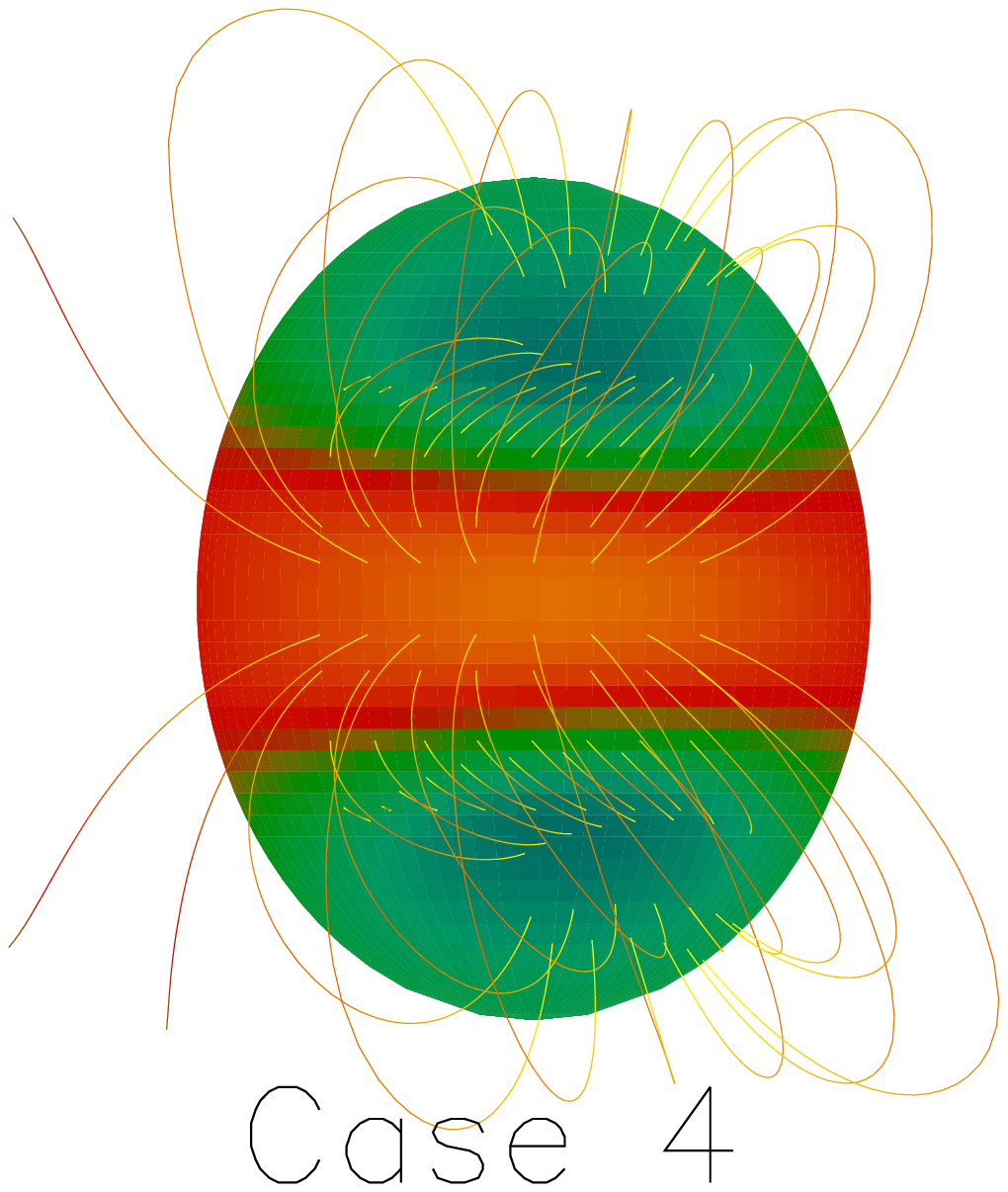}}
\caption{The figure shows the original reference field, a global potential field and
the results of a nonlinear force-free reconstruction with different boundary
conditions (Case 1-4, see text). The color coding shows $B_r$ on the photosphere and
the disk center corresponds to $180^\circ$ longitude.}
 \label{fig2}
\end{figure}

\begin{figure}
\includegraphics[height=10cm,width=12cm]{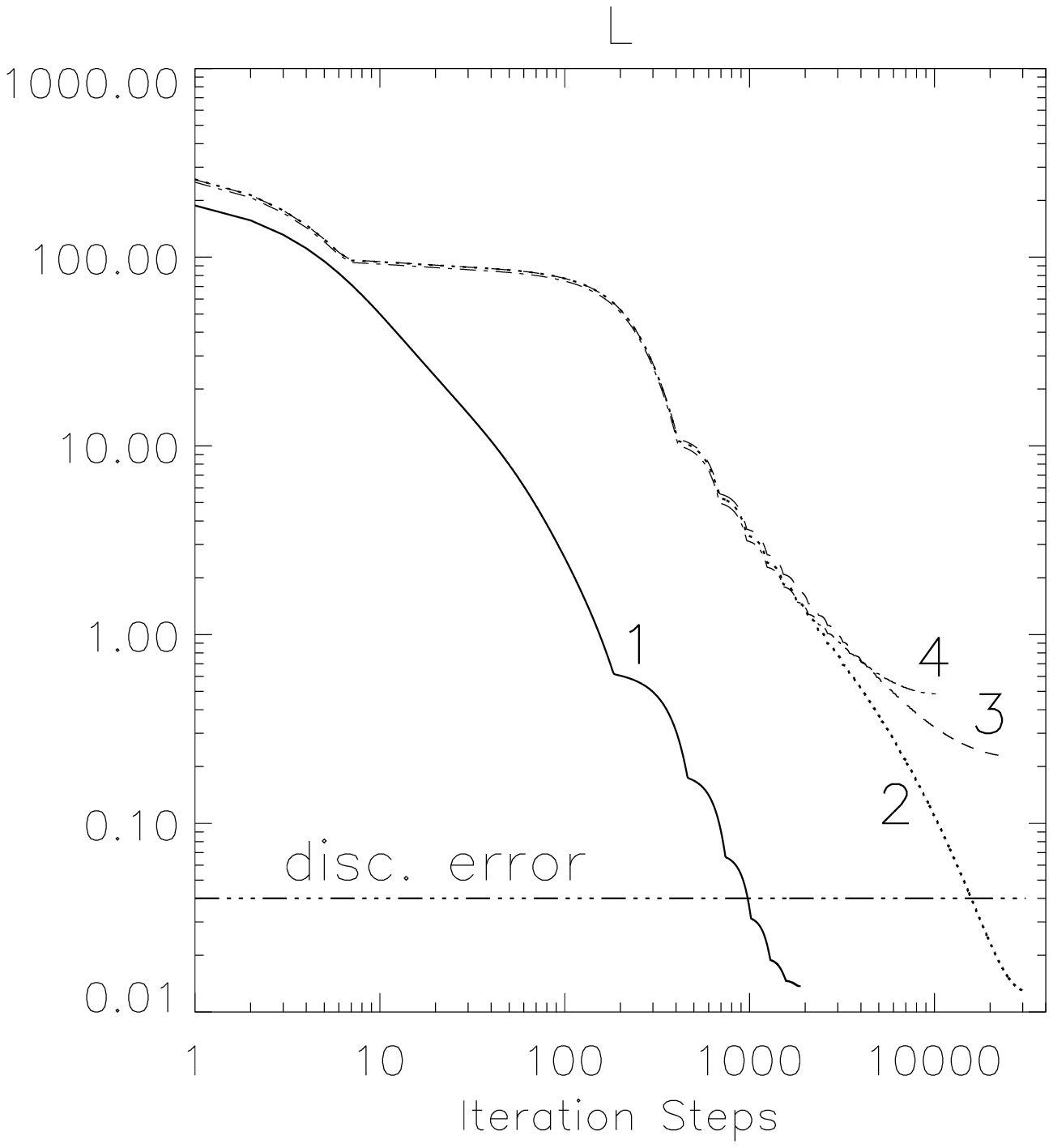}
\includegraphics[height=10cm,width=12cm]{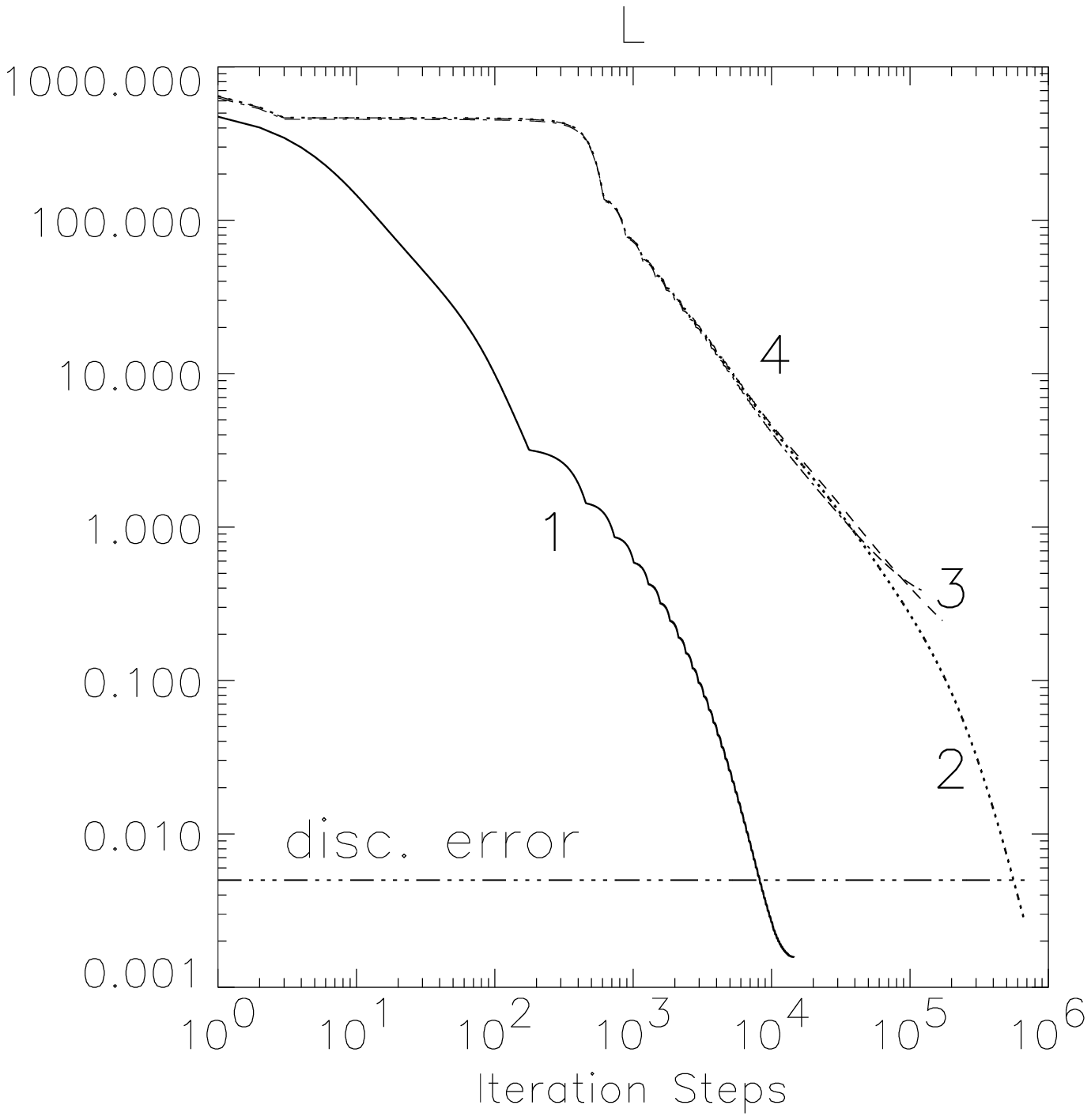}
\caption{Evolution of $L$ (as defined in \ref{defL1})  during the optimization
process. The solid line corresponds to Case 1, the dotted line to Case 2, the dashed
line to Case 3 and the dash-dotted line to Case 4. The horizontal dash-double-dotted
line marks the discretization error of the original semi analytic solution. The top
panel corresponds to the low resolution $(20 \times 40\times 80)$ and the bottom
 panel to the high resolution $(40 \times 80\times 160)$ computation.}
 \label{fig3}
\end{figure}
\subsection{Semi-Analytic Reference Field}
 We test our newly developed code with the help of a
 known nonlinear force-free field model developed by
 \inlinecite{low:etal90}.
 The authors solved the Grad-Shafranov equation for axisymmetric
 force-free fields in spherical coordinates $r$, $\theta$, $\phi$.
 The magnetic field can be written in the form
\begin{equation}
{\bf B} = \frac{1}{r \sin\theta} \; \left( \frac{1}{r} \, \frac{\partial A}{\partial
\theta} {\bf e}_r- \frac{\partial A}{\partial r} {\bf e}_{\theta} + Q \, {\bf
e}_{\phi} \right) \label{lowloub}
\end{equation}
where $A$ is the flux function, and $Q$ represents the $\phi$-component of ${\bf
B}$, depending only on $A$. The flux function $A$ satisfies the Grad-Shafranov
equation
\begin{equation}
 \frac{\partial^2 A}{\partial r^2}+\frac{1-\mu^2}{r^2} \,
\frac{\partial^2 A}{\partial \mu^2}+Q \; \frac{d \, Q}{d \, A} =0
\end{equation}
where $\mu=\cos\theta$. \inlinecite{low:etal90} derive solutions for
\begin{equation}
\frac{d  Q}{d  A} = \alpha= \mbox{constant}
\end{equation}
by looking for separable solutions of the form
\begin{equation}
A(r,\theta)=\frac{P(\mu)}{r^n}.
\end{equation}
The solutions are axisymmetric in spherical coordinates with a point source at the
origin. They have become a kind of standard test for nonlinear force-free
extrapolation codes
 \cite{amari:etal99,wheatland:etal00,wiegelmann:etal03,yan:etal06,amari:etal06,inhester:etal06,schrijver:etal06}
  in Cartesian geometry because the symmetry in the solution
is no longer obvious after a translation which places the point source outside the
computational domain and a rotation of the symmetry axis with respect to the
Cartesian coordinate axis.

 Here we use the Low and Lou solution in spherical coordinates. The original
 equilibrium is invariant in $\phi$,
 but we can produce a 3-D looking configuration by placing the origin of the
 solution with $1/4$ solar radius offset to the sun center. The corresponding
 configuration is not symmetric
 in $\phi$ anymore with respect to the solar surface as seen in the synoptic map
 in Figure \ref{fig1} which shows $B_r$ on the photosphere. Let us remark that
 we use the solution for the purpose of testing our code only and the equilibrium is not
 assumed to be a realistic model for the global coronal magnetic field. We do
 the test runs on spherical grids $(r, \theta, \phi)$ of $20 \times 40 \times 80$ and
 $40 \times 80 \times 160$ grid points.
\subsection{Boundary Conditions}
\label{boundary}
\begin{itemize}
 \item Case 1: Boundary specified on the photosphere, source surface and in
co-latitude at $\theta=30^\circ$ and $\theta=150^\circ$. Optimization restricted in co-latitude to
$30^\circ < \theta < 150^\circ$.
 \item Case 2: Boundary specified on photosphere,
 the source surface and in co-latitude at $\theta=\frac{d \theta}{2}$
 and $\theta=180-\frac{d \theta}{2}$.
 \item Case 3: Boundary specified on photosphere  and
 on the source surface.
 \item Case 4: Boundary specified only on the photosphere.
This is the realistic Case for real data, where measurements are only available on
the photosphere. On the source surface the magnetic field is chosen from the initial
 potential field.
\end{itemize}
 For the computations done here we use a grid resolution of
 $d \theta= d \phi=4.5^\circ$ for the low and $d \theta=d \phi= 2.25^\circ$ for
 the high resolution test Case
 \footnote{If the angle is given in radian we have even
 $d r=d \theta = d \phi$ for the test cases. The code allows, however, also
 the use of $d r \not=d \theta \not= d \phi$.}.
 This means that the full spherical
 extrapolation cases (Case 2- Case 4) are limited in co-latitude
 by $2.25^\circ \leq \theta \leq 177.75^\circ$ and $1.125^\circ \leq \theta \leq 178.875^\circ$
 for the low and high resolution computations, respectively.
 As the initial state we compute a source surface potential field
 in our computational domain. The source surface is a spherical shell where
 all field lines are assumed to become radial \cite{schatten69}.
 We locate the source surface at $1+\frac{\pi}{2} \approx 2.57 R_{\rm s}$ which
 is the outer radial boundary of our physical domain. The finite differences
 in $\phi$ are cyclic. For Case 1 and 2 boundary values in $\theta$ are specified
 and for Case 3 and 4 we interpolate the values at the poles.

\section{Figures of Merit}
\label{merit}
 In Table \ref{table1} we provide some quantitative measures to
 rate the quality of our reconstruction.
 Column 1 names the corresponding test case.
 Column 2-4 show, how well the force and solenoidal condition are fulfilled,
 where column 1 contains the value of the functional $L$ as defined in
 Equation (\ref{defL1}) and $L_1$ and $L_2$ in column 3 and 4 correspond
 to the first (force-free) and second (solenoidal free) part of $L$.
 The evolution of the functional $L$ during the optimization process
 is shown in Figure \ref{fig3}.
 Column 5 contains the $L_{\infty}$ norm of the divergence of the magnetic
 field
\begin{equation}
\parallel \nabla \cdot {\bf B} \parallel_{\infty}=
\sup_{{\bf x} \in V} |\nabla \cdot {\bf B}|
\end{equation}
and column 6 the $L_{\infty}$ norm of the Lorentz force of the magnetic
 field
\begin{equation}
\parallel {\bf j } \times {\bf B} \parallel_{\infty}=
\sup_{{\bf x} \in V} |{\bf j } \times {\bf B} | .
\end{equation}

 The next five columns of Table \ref{table1} contain different measures
 which compare
 our reconstructed field with the semi-analytic reference field. These
 measures have been introduced by \inlinecite{schrijver:etal06} to
 compare a vector field ${\bf b}$ with a reference field ${\bf B}$.

\begin{itemize}
\item Column 7: Vector correlation:
\begin{equation}
C_{\rm vec}=  \sum_i {\bf B_i} \cdot {\bf b_i}/ \left( \sum_i |{\bf B_i}|^2 \sum_i
|{\bf b_i}|^2 \right)^{1/2},
\end{equation}
\item Column 8: Cauchy-Schwarz inequality
\begin{equation}
C_{\rm CS} = \frac{1}{N} \sum_i \frac{{\bf B_i} \cdot {\bf b_i}} {|{\bf B_i}||{\bf
b_i}|},
\end{equation}
where $N$ is the number of vectors in the field.
\item Column 9: Normalized vector error
\begin{equation}
E_{\rm N} = \sum_i |{\bf b_i}-{\bf B_i}|/ \sum_i |{\bf B_i}|,
\end{equation}
\item Column 10:  Mean relative vector error
\begin{equation}
E_{\rm M} = \frac{1}{N} \sum_i \frac{|{\bf b_i}-{\bf B_i}|}{|{\bf B_i}|}.
\end{equation}
\item Column 11: Total magnetic energy of the reconstructed field normalized with the
 energy of the input field
\begin{equation}
\epsilon = \frac{\sum_i |{\bf b_i}|^2}{\sum_i |{\bf B_i}|^2}.
\end{equation}
 \end{itemize}
The two vector fields agree perfectly if $C_{\rm vec},C_{\rm CS}$ and $\epsilon$ are
unity and if $E_{\rm N}$ and $E_{\rm M}$ are zero. Column 12 contains the number of
iteration steps until convergence and column 13 shows the computing time on 4
processors.

\begin{table}
\caption{}
The table rates the quality of our reconstructions with several figures of merit as explained \\
in Section \ref{merit}. We compute the figures for the whole sphere. For Case 1, where the \\
computations have been limited in latitude, the reference field was specified in the cones \\
$\theta < 30^\circ$ and $\theta > 150^\circ$. \\
\label{table1}
\hspace*{-3.5cm}
\begin{tabular}{l|lllcc|lllll|rr}     
\hline Model & $L$&$L_1$&$L_2$&
$\parallel \nabla \cdot {\bf B} \parallel_{\infty}$&
$\parallel {\bf j } \times {\bf B} \parallel_{\infty} $ &
$C_{\rm vec}$&$C_{\rm CS}$&$E_{N}$&$E_{M}$&$\epsilon$ &Steps& Time \\
\hline
&\multicolumn{4}{l}{Spherical grid $20 \times 40 \times 80$} &&&&&&&& \\
Original &$0.04$&$0.03$&$0.01$&$0.48$&$1.37$& $ 1$&$ 1$&$ 0$&$ 0$&$ 1$&$ $&$ $ \\
Potential &$0.19$&$0.006$&$0.13$&$4.61$&$0.90$&$0.66$&$0.77$&$0.71$&$0.68$&$0.75$&& \\
Case 1 &$0.014$&$0.010$&$0.004$&$0.42$&$0.64$&$0.9998$&$0.9995$&$0.012$&$0.016$&$1.008$&$1889$& 2 min \\
Case 2 &$0.013$&$0.010$&$0.003$&$0.42$&$0.64$&$0.9998$&$0.9992$&$0.016$&$0.023$&$1.007$&$31129$& 23 min \\
Case 3 &$0.23$&$0.20$&$0.03$&$3.00$&$11.30$&$0.998$&$0.998$&$0.04$&$0.05$&$0.99$&$23824$& 18 min \\
Case 4 &$0.48$&$0.34$&$0.14$&$3.11$&$11.30$&$0.993$&$0.95$&$0.14$&$0.29$&$0.95$&$10167$&  7 min \\
\hline
&\multicolumn{4}{l}{Spherical grid $40 \times 80 \times 160$} &&&&&&&& \\
Original &$0.005$&$0.003$&$0.002$&$0.38$&$0.71$&$ 1$&$ 1$&$ 0$&$ 0$&$ 1$&$ $&$ $ \\
Potential &$0.30$&$0.0003$&$0.30$&$0.44$&$0.23$&$0.67$&$0.77$&$0.70$&$0.67$&$0.75$&$ $&$ $ \\
Case 1 &$0.0016$&$0.0010$&$0.0006$&$0.38$&$0.32$&$0.99998$&$0.99989$&$0.004$&$0.007$&$1.002$&$14522$& 1h 26min \\
Case 2 &$0.0027$&$0.0020$&$0.0007$&$0.40$&$0.53$&$0.99992$&$0.9995$&$0.011$&$0.019$&$0.9989$&$672281$&65h 19min \\
Case 3 &$0.24$&$0.20$&$0.04$&$6.63$&$22.47$&$0.996$&$0.99$&$0.086$&$0.12$&$0.96$&$171143$ & 16h 57min\\
Case 4 &$0.39$&$0.27$&$0.12$&$6.60$&$22.47$&$0.99$&$0.93$&$0.17$&$0.32$&$0.92$&$120742$ & 12h 00min\\
\hline
\end{tabular}
\end{table}
\section{Results}
\label{results}
\subsection{Qualitative Comparison}
 In Figure \ref{fig2} we compare magnetic field line plots of
 the original model field (Original) with a corresponding potential
 field (Potential) and nonlinear force-free reconstructions with different
 boundary conditions (Case 1 - Case 4). The colour coding shows the
 radial magnetic field on the photosphere, as also shown in the
 synoptic map in Figure \ref{fig1}. Carrington longitude $180^\circ$
 corresponds to the disk center in Figure \ref{fig2}. The images
 show the results of the computation on the $20 \times 40 \times 80$ grid.

 A comparison of the original reference field (Original in Figure \ref{fig1})
 with our nonlinear force-free reconstructions (Cases 1-4) shows that
 the magnetic field line plots agree with the original for Case 1-3
 within the plotting precision. Case 4 shows some deviations from the original,
 but the reconstructed field lines are much closer to the reference field
 than to the initial potential field (Potential). We cannot expect a perfect
 agreement with the reference field for Case 4, because here the outer radial
 boundary conditions are taken from the initial potential field model, which
 are different from the outer boundary of the reference field. It is well known
 from computations in Cartesian geometry that the lateral and top boundaries
 influence the solution in the box \footnote{In the Cartesian case this effect
 can be reduced by choosing a well isolated active region surrounded by a sufficiently
 large area with low magnetic field, see e.g., test Case II in \inlinecite{schrijver:etal06}.}.

\subsection{Quantitative Comparison}
 The visual inspection of Figure \ref{fig2} is supported by the quantitative
 criteria shown in Table \ref{table1}. For Cases 1 and 2 (where the boundary conditions
 have been specified on the photosphere, the source surface and in latitude)
 the formal force-free criteria $(L, L_1, L_2)$ are smaller
 than the discretization error of the analytic solution and the
 comparison metrics show an almost perfect agreement with the reference field.
 For Case 3  we still find a very good agreement between the
 reference field and our reconstruction and for Case 4
 the agreement is still reasonable. We are in particular able to
 compute the magnetic energy content of the coronal magnetic field approximately
 correct. A potential field reconstruction does obviously not
 agree with the reference field. The figures of merit show that
 the potential field is far away from the true solutions and contains only
 $75 \%$ of the magnetic energy.

 For a practical use of any numerical scheme the computing time certainly matters.
 As seen in the last column of Table \ref{table1} the computing time is quite
 fast if we restrict the computational domain to low co-latitudes
 ($30^\circ < \theta < 150^\circ, \;$ Case 1) but increases significantly for
 full sphere computations. The time needed for one iteration time step is
 approximately constant ($0.04{\rm s}$ on the low and $0.36{\rm s}$ on the high resolution grid.),
 but the number of iteration steps needed until convergence increases by more
 than one order of magnitude if high co-latitude regions are included in the optimization.
 The reason is that the iteration time step $dt$, which adjusts automatically in our code,
 becomes much smaller and consequently the convergence speed becomes slow. This is well visible
 in Figure \ref{fig3} which shows the evolution of the functional $L$ with the number of
 iteration steps. For the low latitude computations (Case 1, solid line in Figure \ref{fig3})
 the functional $L$ is much steeper than for the full sphere computations (Cases 2-4, dotted,dashed, dash-dotted lines).
 The slow convergence of the full sun computations is directly related to the smaller time step and
 the time step itself is restricted by the physical grid resolution
  \footnote{This condition is similar to the CFL-condition for time dependent problems.}, which becomes very small close
 to the poles. We will address this point in the discussion (Section \ref{conclusions}) and outline
 possible solutions to overcome these difficulties.

 Some insights regarding the performance of our newly developed code might be
 given by a comparison of the figures in Table \ref{table1} with the
 results of computations in Cartesian coordinates, as shown in
 \inlinecite{schrijver:etal06} Table I, where row (b) contains the results
 of our Cartesian optimization code.
 The upper part of Schrijver's
 Table I corresponds to a test Case where all six boundaries of
 a Cartesian box have been specified (Schrijver's Case I, which is somewhat
 similar to our Cases 1-3) and the lower part of \inlinecite{schrijver:etal06} Table I
 corresponds to a case where only the bottom boundary of a Cartesian box has
 been specified (Schrijver's Case II, which is equivalent to our Case 4 in
 spherical geometry).
 Both for Cartesian and spherical computations the
 correspondence with the original field is reduced
 if the boundary conditions are only specified
 on the photosphere. For real observed vector magnetograms we certainly
 have only photospheric data and it is therefore important to get a
 reasonable nonlinear force-free magnetic field reconstruction for
 this case. The errors of the different comparison metrics are
 still small and in a comparable range for Cartesian and spherical
 computations. The vector correlation is better than $99 \%$ and
 we got the magnetic energy correct within a few percent for
 all examples investigated in this paper.

\section{Discussion and Outlook}
\label{conclusions}
 Within this work we developed a code for the nonlinear force-free
 computation of coronal magnetic fields in spherical geometry.
 The method is based on an optimization principle. We tested the
 performance of the newly developed code with the help of a semi-analytic
 reference field. We find that the
 spherical optimization method works well and the accuracy of the
 reconstructed magnetic field configuration is comparable with
 the performance of a corresponding code in Cartesian geometry.
 The computation is reasonably fast if we limit the computation
 to low latitude regions,
 but becomes significantly slower if polar regions are included.
 Such a behaviour is well known for computations in spherical coordinates,
 see e.g. \inlinecite{kageyama:etal04}. The reason is that the
 physical grid converges to the poles. A fair approximation for
 the iteration time step is $dt \propto \Delta^2$, where $\Delta$ is
 the physical grid resolution. As the finest grid resolution restricts
 the time step, it becomes much smaller if high altitude regions
 (where $\Delta$ becomes very small) are included
 in the optimization. Our code has an automatic step size control and
 we find that $dt$ automatically adjusts to much smaller values for
 full sphere computations.

 A possible solution is the use of the so called Yin-Yang grid, as developed by
 \inlinecite{kageyama:etal04} and as used, e.g., for Earth mantle convection by
\inlinecite{yoshida:etal04}. The Yin-Yang grid is composed of two identical
complimentary grids, which partly overlap and together cover the solar surface with
a quasi-uniform grid spacing (see \inlinecite{kageyama:etal04} for details).
Naturally the Yin-Yang grid has no poles at all and a considerably large iteration
step can be used. There is certainly a numerical overhead for transformations
between the complimentary grids, but the limitations regarding the time step in
polar regions for traditional spherical grids are more time demanding. For solar
applications one has to consider, however, that photospheric magnetic field
measurements are currently less accurate close to the poles. It might therefore be
acceptable to restrict a nonlinear force-free computations onto equatorial regions,
say between about $30^\circ$ and $150^\circ$ latitude \footnote{Alternatively one could use a
larger grid spacing for the computation of a force-free field in polar regions}. For
these low latitude regions the spherical optimization code as described here is
reasonably fast. For application to observed vector magnetograms one has to
consider, that the measured  magnetic field is not necessary force-free in the
photosphere and in particular the transverse components of the magnetic field
vector contain much noise. These problems are, however, also present for force-free
extrapolations in Cartesian geometry and a preprocessing of the photospheric vector
magnetograms as described in \inlinecite{wiegelmann:etal06} helps to overcome these
difficulties.

  The code solves the nonlinear force-free equations in the bounded domain
    between $1 Rs$ and the source surface at $2.57 Rs$. The outer boundary is kept
    fixed from the initial potential field. All current carrying field lines
    have to close inside the volume. The domain outside $2.57 Rs$ is not included
    in the model, because the force-free approach is not justified anymore here.
 A further step towards a consistent modelling of the solar corona would be the
 inclusion of non-magnetic forces, like plasma pressure, gravity and the dynamic
 pressure of the solar wind.
 This is in particular useful for several solar radii long structures
 like helmet streamer,
 where the plasma $\beta$ becomes finite
 (see e.g., \inlinecite{guo:etal98} and \inlinecite{wiegelmann:etal98}).
 \inlinecite{neukirch95} found a special class of magnetohydrostatic
 solutions which are separable in spherical coordinates.
 \inlinecite{petrie:etal00} extended the linear force-free Green's function
 methods towards the inclusion of non-magnetic forces in Cartesian geometry.
 Both approaches assume a kind of global linear force-free parameter $\alpha$
 for the parallel part of the electric currents. This is certainly a too
 restrictive condition to include the detailed information provided by
 measured vector magnetograms.
 The optimization method has been generalized for this aim in
 \inlinecite{wiegelmann:etal03a} and implemented in Cartesian geometry.
 A corresponding implementation into spherical geometry is straightforward,
 but does certainly require additional observational constraints, e.g.,
 the tomographically reconstructed 3D coronal density distribution.

%
\begin{acknowledgements}
 This work was supported by DLR-grant 50 OC 0501. The author thanks
 Bernd Inhester for useful comments and acknowledge inspiriting
 discussions on three workshops organized by Karel Schrijver
 (NLFFF-consortium) in Palo Alto between 2004 and 2006.
We thank the referee, Tahar Amari, for useful remarks to improve this paper.
\end{acknowledgements}


\begin{thebibliography}{}

\bibitem[\protect\citeauthoryear{Amari, Boulmezaoud, and Aly}{2006}]{amari:etal06}
Amari, T., Boulmezaoud, T.~Z., and Aly, J.~J.: 2006,
{\it Astron. Astrophys.} {\bf 446}, 691.

\bibitem[\protect\citeauthoryear{Amari, Boulmezaoud, and Mikic}{1999}]{amari:etal99}
Amari, T., Boulmezaoud, T.~Z., and Mikic, Z.: 1999,
{\it Astron. Astrophys.} {\bf 350}, 1051.

\bibitem[\protect\citeauthoryear{Amari et al.}{1997}]{amari:etal97}
Amari, T., Aly, J.~J., Luciani, J.~F., Boulmezaoud, T.~Z., and Mikic, Z.: 1997, {\it Sol. Phys.} {\bf 174}, 129.

\bibitem[\protect\citeauthoryear{Bleybel et al.}{2002}]{bleybel:etal02}
Bleybel, A., Amari T., van Driel-Gesztelyi, L., and Leka, K.~D.: 2002, {\it Astron. Astrophys.}, {\bf 395}, 685.

\bibitem[\protect\citeauthoryear{{Carcedo} et al.}{2003}]{carcedo:etal03}
Carcedo, L., Brown, D.~S., Hood, A.~W., Neukirch, T., and Wiegelmann, T.: 2003, {\it Sol. Phys.} {\bf 218}, 29.

\bibitem[\protect\citeauthoryear{{Chiu} and {Hilton}}{1977}]{chiu:etal77}
Chiu, Y.~T. and Hilton, H.~H.: 1977, {\it Astrophys. J.} {\bf 212}, 873.

\bibitem[\protect\citeauthoryear{Cuperman, Demoulin, and Semel}{1991}]{cuperman:etal91}
Cuperman, S., Demoulin, P., and Semel M.: 1991,
{\it Astron. Astrophys.} {\bf 245}, 285.

\bibitem[\protect\citeauthoryear{Demoulin, Cuperman, and Semel}{1992}]{demoulin:etal92}
Demoulin, P., Cuperman, S., and Semel M.: 1992,
{\it Astron. Astrophys.} {\bf 263}, 351.

\bibitem[\protect\citeauthoryear{{Grad} and {Rubin}}{1958}]{grad:etal58}
Grad, H. and Rubin, H.: 1958, in
 {\it Proceedings of the 2nd International Conference on Peaceful Uses of Atomic Energy} {\bf 31}, United Nations, Geneva, 190.

\bibitem[\protect\citeauthoryear{{Guo} and {Wu}}{1998}]{guo:etal98}
Guo, W.~P. and Wu, S.~T.: 1998,
{\it Astrophys. J.} {\bf 494}, 419.

\bibitem[\protect\citeauthoryear{{Inhester} and
  {Wiegelmann}}{2006}]{inhester:etal06}
Inhester, B. and Wiegelmann, T.: 2006,
{\it Sol. Phys.} {\bf 235}, 201.

\bibitem[\protect\citeauthoryear{{Kageyama} and {Sato}}{2004}]{kageyama:etal04}
Kageyama, A. and Sato, T.: 2004,
{\it Geochemistry, Geophysics, Geosystems} {\bf 5}, 9005.

\bibitem[\protect\citeauthoryear{{Lin} et~al.}{2004}]{lin:etal04}
Lin, H., Kuhn, J.~R., and Coulter, R.: 2004,
{\it Astrophys. J.} {\bf 613}, L177.

\bibitem[\protect\citeauthoryear{{Low} and {Lou}}{1990}]{low:etal90}
Low, B.~C. and Lou, Y.~Q.: 1990,
{\it Astrophys. J.} {\bf 352}, 343.

\bibitem[\protect\citeauthoryear{{Marsch} et al.}{2004}]{marsch:etal04}
Marsch, E., Wiegelmann, T., and Xia, L.~D.: 2004,
{\it Astron. Astrophys.} {\bf 428}, 629.

\bibitem[\protect\citeauthoryear{Mikic and McClymont}{1994}]{mikic:etal94}
Mikic, Z. and McClymont, A.~N:, 1994, in
K.~S. Balasubramaniam and G.~W. Simon (eds.),
{\it Solar Active Region Evolution: Comparing Models with Observations}, {\it Astron. Soc. Pacific Conf. Ser.} {\bf 68}, 225.

\bibitem[\protect\citeauthoryear{{Neugebauer} et al.}{2002}]{neugebauer:etal02}
Neugebauer, M., Liewer, P.~C., Smith, E.~J., Skoug, R.~M., and Zurbuchen, T.~H.: 2002, {\it J. Geophys. Res.} {\bf 107}(13), 1.


\bibitem[\protect\citeauthoryear{{Neukirch}}{1995}]{neukirch95}
Neukirch, T.: 1995,
{\it Astron. Astrophys.} {\bf 301}, 628.

\bibitem[\protect\citeauthoryear{{Petrie} and {Neukirch}}{2000}]{petrie:etal00}
Petrie, G.~J.~D. and Neukirch, T.: 2000,
{\it Astron. Astrophys.} {\bf 356}, 735.

\bibitem[\protect\citeauthoryear{R{\'e}gnier and Amari}{2004}]{regnier:etal04}
R{\'e}gnier, S. and Amari, T.: 2004,
{\it Astron. Astrophys.} {\bf 425}, 345.

\bibitem[\protect\citeauthoryear{R{\' e}gnier, Amari, and Kersal{\' e}}{2002}]{regnier:etal02}
R{\' e}gnier, S., Amari, T., and Kersal{\' e}, E.: 2002,
{\it Astron. Astrophys.} {\bf 392}, 1119.

\bibitem[\protect\citeauthoryear{{Roumeliotis}}{1996}]{roumeliotis96}
Roumeliotis, G.: 1996,
{\it Astrophys. J.} {\bf 473}, 1095.

\bibitem[\protect\citeauthoryear{{Sakurai}}{1981}]{sakurai81}
Sakurai, T.: 1981,
{\it Solar Phys.} {\bf 69}, 343.

\bibitem[\protect\citeauthoryear{{Schatten} et al.}{1969}]{schatten69}
Schatten, K.~H., Wilcox, J.~M., and Ness, N.~F.: 1969,
{\it Solar Phys.} {\bf 6}, 442.

\bibitem[\protect\citeauthoryear{{Schrijver} and
  {Derosa}}{2003}]{schrijver:etal03}
Schrijver, C.~J. and Derosa, M.~L.: 2003,
{\it Solar Phys.} {\bf 212}, 165.

\bibitem[\protect\citeauthoryear{{Schrijver} et al.}{2006}]{schrijver:etal06}
Schrijver, C.~J., Derosa, M.~L., Metcalf, T.~R., Liu, Y., McTiernan, J.,
R{\'e}gnier, S., Valori, G., Wheatland, M.~S., and Wiegelmann, T.: 2006, {\it Solar Phys.} {\bf 235}, 161.

\bibitem[\protect\citeauthoryear{{Seehafer}}{1978}]{seehafer78}
Seehafer, N.: 1978,
{\it Solar Phys.} {\bf 58}, 215.

\bibitem[\protect\citeauthoryear{{Solanki} et al.}{2003}]{solanki:etal03}
Solanki, S.~K., Lagg, A., Woch, J., Krupp, N., and Collados, M.: 2003, {\it Nature} {\bf 425}, 692.

\bibitem[\protect\citeauthoryear{{Valori} et al.}{2005}]{valori:etal05}
Valori, G., Kliem, B., Keppens, R.: 2005,
{\it Astron. Astrophys.} {\bf 433}, 335.

\bibitem[\protect\citeauthoryear{{Wheatland}}{2004}]{wheatland04}
Wheatland, M.~S.: 2004,
{\it Solar Phys.} {\bf 222}, 247.

\bibitem[\protect\citeauthoryear{{Wheatland} et al.}{2000}]{wheatland:etal00}
Wheatland, M.~S., Sturrock, P.~A., and Roumeliotis, G.: 2000, {\it Astrophys. J.} {\bf 540}, 1150.

\bibitem[\protect\citeauthoryear{Wiegelmann and Inhester}{2006}]{wiegelmann:etal06a}
Wiegelmann, T. and Inhester, B.: 2006,
{\it Solar Phys.} {\bf 236}, 25.

\bibitem[\protect\citeauthoryear{Wiegelmann, Inhester and Sakurai}{2006}]{wiegelmann:etal06}
Wiegelmann, T., Inhester, B., and Sakurai, T.: 2006,
{\it Solar Phys.} {\bf 233}, 215.

\bibitem[\protect\citeauthoryear{{Wiegelmann} et al.}{2005}]{wiegelmann:etal05}
Wiegelmann, T., Lagg, A., Solanki, S.~K., Inhester, B., and Woch, J.: 2005, {\it Astron. Astrophys.} {\bf 433}, 701.

\bibitem[\protect\citeauthoryear{Wiegelmann et al.}{2005a}]{wiegelmann:etal05a}
Wiegelmann, T., Inhester, B., Lagg, A., and Solanki, S.~K.: 2005, {\it Solar Phys.} {\bf 228}, 67.

\bibitem[\protect\citeauthoryear{{Wiegelmann}}{2004}]{wiegelmann04}
Wiegelmann, T.: 2004,
{\it Solar Phys.} {\bf 219}, 87.

\bibitem[\protect\citeauthoryear{{Wiegelmann} and
  {Inhester}}{2003}]{wiegelmann:etal03a}
Wiegelmann, T. and Inhester, B.: 2003,
{\it Solar Phys.} {\bf 214}, 287.

\bibitem[\protect\citeauthoryear{{Wiegelmann} and
  {Neukirch}}{2003}]{wiegelmann:etal03}
Wiegelmann, T. and Neukirch, T.: 2003,
{\it Nonlinear Processes in Geophysics} {\bf 10}, 313.

\bibitem[\protect\citeauthoryear{{Wiegelmann} et al.}{1998}]{wiegelmann:etal98}
Wiegelmann, T., Schindler, K., and Neukirch, T.: 1998,
{\it Solar Phys.} {\bf 180}, 439.

\bibitem[\protect\citeauthoryear{{Wu} et al.}{1990}]{wu:etal90}
Wu, S.~T., Sun, M.~T., Chang, H.~M., Hagyard, M.~J., and Gary, G.~A.: 1990, {\it Astrophys. J.} {\bf 362}, 698.

\bibitem[\protect\citeauthoryear{{Yan} and {Li}}{2006}]{yan:etal06}
Yan, Y. and Li, Z.: 2006,
{\it Astrophys. J.} {\bf 638}, 1162.

\bibitem[\protect\citeauthoryear{{Yan} and {Sakurai}}{2000}]{yan:etal00}
Yan, Y. and Sakurai, T.: 2000,
{\it Solar Phys.} {\bf 195}, 89.

\bibitem[\protect\citeauthoryear{{Yoshida} and
  {Kageyama}}{2004}]{yoshida:etal04}
Yoshida, M. and Kageyama, A.: 2004,
{\it Geophys. Res. Lett.} {\bf 31}, 12609.

\end{thebibliography}

\end{article}
\end{document}